\documentclass[12pt]{article}

\begin{document}
\pagenumbering{roman}
\def\Barcelo{Barcel\'o}
\title{\bf Twilight for the energy conditions?}
\author{
Carlos {\Barcelo}$^{*}$ \\[2mm]
{\small \it 
Institute of Cosmology and Gravitation,  University of Portsmouth}
\\ 
{\small \it Portsmouth PO1 2EG, Britain.}
\\[12pt]
Matt Visser$^{\dagger}$ \\[2mm]
{\small \it 
Physics Department, Washington University,}
\\ 
{\small \it 
Saint Louis, MO 63130-4899, USA.}
\\[12pt]
} 
\date{{\small 25 March 2002; \LaTeX-ed \today}}
\maketitle
\vfill
\hrule
\bigskip
\centerline{$^{*}${\sf carlos.barcelo@port.ac.uk}}
\centerline{$^{\dagger}${\sf visser@kiwi.wustl.edu}; 
http://www.physics.wustl.edu/\~{}visser}
\bigskip
\hrule
\bigskip
\noindent
{$^{\dagger}$After 1 July 2002: 
\\
School of Mathematics and Computer Science,
Victoria University, 
\\
PO Box 600, Wellington, New Zealand;
\\
{\sf
matt.visser@mcs.vuw.ac.nz}; http://www.mcs.vuw.ac.nz/\~{}visser
\bigskip
\hrule
\bigskip
\noindent
This essay received an ``honourable mention'' in the 2002 essay
competition sponsored by the Gravity Research Foundation.
\bigskip

\centerline{\underline{Archive:} {\sf gr-qc/0205066}}

\bigskip
\hrule
\clearpage
\markright{Twilight for the energy conditions?\hfil }
\pagestyle{myheadings}

\null
\vskip 2cm

\begin{abstract}

\bigskip

The tension, if not outright inconsistency, between quantum physics
and general relativity is one of the great problems facing physics at
the turn of the millennium. Most often, the problems arising in
merging Einstein gravity and quantum physics are viewed as Planck
scale issues ($10^{19}$~GeV, $10^{-34}$~m, $10^{-45}$~s), and so
safely beyond the reach of experiment. However, over the last few
years it has become increasingly obvious that the difficulties are
more widespread: There are already serious problems of deep and
fundamental principle at the semi-classical level, and worse, certain
classical systems (inspired by quantum physics, but in no sense
quantum themselves) exhibit seriously pathological behaviour.  One
manifestation of these pathologies is in the so-called ``energy
conditions'' of general relativity.  Patching things up in the gravity
sector opens gaping holes elsewhere; and some ``fixes'' are more
radical than the problems they are supposed to cure.

\vspace*{5mm}
\end{abstract}

\vfill

\clearpage

\def\eg{{\em e.g.}}
\def\S{{\mathcal S}}
\def\I{{\mathcal I}}
\def\L{{\mathcal L}}
\def\eff{{\mathrm{eff}}}
\def\Newton{{\mathrm{Newton}}}
\def\bulk{{\mathrm{bulk}}}
\def\matter{{\mathrm{matter}}}
\def\tr{{\mathrm{tr}}}
\def\normal{{\mathrm{normal}}}
\def\half{{1\over2}}
\def\etc{{\emph{etc}}}
\def\HRULE{{\bigskip\hrule\bigskip}}



\setcounter{page}{1}
\pagenumbering{arabic}

\section{Energy conditions of General Relativity}

Even if you restrict attention to the purely classical regime,
Einstein gravity (general relativity) is a tremendously complex
theory. In the field equations $G^{\mu\nu} = 8\pi \; G\; T^{\mu\nu}$,
the left-hand-side, the Einstein tensor $G^{\mu\nu}$, is complicated
enough by itself. But it is at least a universal function of the
spacetime geometry. In contrast the right-hand-side, the stress-energy
tensor $T^{\mu\nu}$, is not universal but instead depends on the
particular type of matter and interactions you choose to insert in
your model. Faced with this situation, you must either resign oneself
to performing an immense catalog of special-case calculations, one
special case for each conceivable matter Lagrangian you can write
down, or try to decide on some generic features that ``all
reasonable'' stress-energy tensors should satisfy, and then try to use
these generic features to develop general theorems concerning the
strong-field behaviour of gravitational fields.

One key generic feature that most matter we run across experimentally
seems to share is that energy densities (almost) always seem to be
positive. The so-called ``energy conditions'' of general
relativity~\cite{energy-conditions} are a variety of different ways of
making this notion of locally positive energy density more
precise. The (pointwise) energy conditions take the form of assertions
that various linear combinations of the components of the
stress-energy tensor should be positive, or at least
non-negative. (See Table I.)  The so-called ``averaged energy
conditions'' are somewhat weaker, they permit localized violations of
the energy conditions, as long as ``on average'' the energy conditions
hold when integrated along null or timelike geodesics.

The variety of energy conditions in use in the relativity community is
driven largely by reverse engineering based on the technical
requirements of how much you have to assume to easily prove the result
you want. By assuming some form of energy condition, some notion of
positivity of the stress-energy tensor, as an input hypothesis, it has
been possible to prove theorems like the singularity theorems
(guaranteeing, under certain circumstances, gravitational collapse
and/or the existence of a big bang singularity), the positive energy
theorem (guaranteeing the mass of a complex gravitating system as seen
from infinity is always positive), the topological censorship theorem
(guaranteeing the non-existence of traversable wormholes) or the
superluminal censorship (limiting the extent to which light cones can
``tip over'' in strong gravitational fields).  Conversely, the
violation of some or all of these energy conditions would point
towards exotic physical possibilities (see \cite{consequences} for
some examples).

Over the years, opinions have changed as to how fundamental some of
the specific energy conditions are. One particular energy condition
(the trace energy condition, TEC) has now been completely abandoned
and forgotten. The TEC was the assertion that the trace of the
stress-energy tensor should always be negative (or positive depending
on metric conventions), and was popular for a while during the
1960's. However, once it was realized that stiff equations of state,
such as those appropriate for neutron stars, violate the TEC this
energy condition fell into disfavour. It has now been completely
abandoned and is no longer cited in the literature --- we mention it
here as a concrete example of an energy condition being outright
abandoned.

\bigskip
\centerline{\bf Table I: Pointwise Energy Conditions} 
\bigskip
\hspace{-2cm}
\begin{tabular}[l]{|c|c|c|c|}
\hline
Name & Abbreviation & Definition & Current status \\
\hline
\hline
Trace energy condition & TEC & $\rho - 3 p \geq 0$ & forgotten \\
Strong energy condition & SEC & $\;\;\rho+3p\geq 0$; \quad $\rho+p \geq 0\;\;$ & 
dead\\
Null energy condition & NEC & $\rho + p \geq 0$ & moribund\\
Weak energy condition & WEC & $\rho\geq 0$; \quad $\rho_p\geq 0$ & moribund \\
Dominant energy condition& DEC& $\rho\geq0$; \quad $p\in[-\rho,+\rho]$ & moribund\\
\hline
\end{tabular}
\bigskip

There is also general agreement that the strong energy condition (SEC)
is dead: (1) The most naive scalar field theory you can write down,
the minimally coupled scalar field, violates the SEC, and indeed
curvature-coupled scalar field theories also violate the SEC; there
are fermionic quantum field theories where interactions engender SEC
violations, and specific models of point-like particles with two-body
interactions that violate the SEC. (2) If you believe in cosmological
inflation, the SEC must be violated during the inflationary epoch, and
the need for this SEC violation is why inflationary models are
typically driven by scalar inflaton fields. (3) If you believe the
recent observational data regarding the accelerating universe, then
the SEC is violated on cosmological scales {\em right now!} (4) Even
if you are somewhat more conservative, and regard the alleged
present-day acceleration of the cosmological expansion as
``unproven'', the tension between the age of the oldest stars and the
measured present-day Hubble parameter makes it very difficult to avoid
the conclusion that the SEC must have been violated in the
cosmologically recent past, sometime between redshift 10 and the
present \cite{Science}.  Under these circumstances it would be rather
quixotic to take the SEC seriously as fundamental physics.

The null, weak, and dominant energy conditions are on the verge of
dying. Specifically: Over the last decade or so it has become
increasingly obvious that there are quantum effects that are capable
of violating {\em all} the energy conditions, even the weakest of the
standard energy conditions.  Despite the fact that they are moribund,
for lack of truly successful replacements, the NEC, WEC, and DEC are
still extensively used in the general relativity community. The
weakest of these is the NEC, and it is in many cases also the easiest
to work with and analyze. The standard wisdom for many years was that
all reasonable forms of matter should at least satisfy the NEC. After
it became clear that the NEC (and even the ANEC) was violated by
quantum effects two main lines of retrenchment developed:

(1) Many researchers simply decided to ignore quantum mechanics,
relying on the classical NEC to prevent grossly weird physics in the
classical regime, and hoping that the long sought for quantum theory
of gravity would eventually deal with the quantum problems. This is
not really a satisfactory response in that NEC violations already show
up in semiclassical quantum gravity (where you quantize the matter
fields and keep gravity classical), and show up at first order in
$\hbar$. Since semiclassical quantum gravity is certainly a good
approximation in our immediate neighbourhood, it is somewhat
disturbing to see widespread (albeit small) violations of the energy
conditions in the here and now. Many experimental physicists and
observational astrophysicists react quite heatedly when the
theoreticians tell them that according to our best calculations there
should be ``negative energy'' (energy densities less than that of the
flat-space Minkowski vacuum) out there in the real universe. However,
to avoid the conclusion that quantum effects can and do lead to
locally negative energy densities, and even violations of the ANEC,
requires truly radical surgery to modern physics, and in particular we
would have to throw away almost all of quantum field theory.

(2) A more nuanced response is based on the Ford--Roman {\em Quantum
Inequalities}~\cite{Ford-Roman}. These inequalities are based on the
fact that while quantum-induced violations of the energy conditions
are widespread they are also {\em small}, and on the observation that
a negative energy in one place and time always seems to be compensated
for (indeed, over-compensated for) by positive energy elsewhere in
spacetime. This is the so-called {\em Quantum Interest
Conjecture}. While the positive pay-back is not enough to prevent
violation of the ANEC (based on averaging the NEC along a null
geodesic) the hope is that it will be possible to prove some improved
type of space-time averaged energy condition from first principles,
and that such a space-time averaged energy condition might be
sufficient to enable us to recover the
singularity/positive-mass/censorship theorems under weaker hypotheses
than currently employed. (Note that this would not eliminate the
possibility of weird geometrical effects in the subatomic realm.)

A fundamental problem for this type of approach that is now becoming
acute is the realization that there are also serious {\em classical}
violations of the energy conditions~\cite{class-violations}.
Recently, it has become clear that there are quite reasonable looking
classical systems, field theories that are compatible with all known
experimental data, and that are in some sense very natural from a
quantum field theory point of view, which violate all the energy
conditions. Because these are now classical violations of the energy
conditions they can be made arbitrarily large, and seem to lead to
rather weird physics. (For instance, it is possible to demonstrate
that Lorentzian-signature traversable wormholes arise as [unstable]
classical solutions of the field equations.) These classical energy
condition violations are due to the behaviour of scalar fields when
coupled to gravity, so let us devote the next section to present some
background on the usefulness and need for scalar field theories in
modern physics.

However, before finishing the section, and for completeness, we would
like to point out another area present-day physics in which one is
confronted with energy condition violations, namely negative tension
braneworlds. If physics is what physicists do, then negative tension
branes are physics---they are common ancillary objects in braneworld
cosmologies based on variants of the Randall--Sundrum construction.
For our present purposes this is important because negative tension
branes provide classical violations of all the energy conditions in
the higher-dimensional spacetime~\cite{surgery}, and they do so in a
way that is completely independent of your opinions concerning scalar
fields.  These classical violations of the energy conditions easily
engender arbitrarily weird physics.

\section{Scalar Fields}

Scalar fields play a somewhat ambiguous role in modern theoretical
physics: on the one hand they provide great toy models, and are from a
theoretician's perspective almost inevitable components of any
reasonable model of empirical reality; on the other hand the direct
experimental/observational evidence is spotty.

The only scalar fields for which we have really direct ``hands-on''
experimental evidence are the scalar mesons (pions $\pi$; kaons $K$;
and their ``charmed'', ``truth'' and ``beauty'' relatives, plus a
whole slew of resonances such as the $\eta$, $f_0$, $\eta'$, $a_0$,
\dots). Not a single one of these particles are fundamental, they are all
quark-antiquark bound states, and while the description in terms of
scalar fields is useful when these systems are probed at low momenta
(as measured in their rest frame) we should certainly not continue to
use the scalar field description once the system is probed with
momenta greater than $\hbar/({\mathrm{bound~state~radius}})$.
Similarly you should not trust the scalar field description if the
energy density in the scalar field exceeds the critical density for
the quark-hadron phase transition. Thus scalar mesons are a mixed bag:
they definitely exist, and we know quite a bit about their properties,
but there are stringent limitations on how far we should trust the
scalar field description.

The next candidate scalar field that is closest to experimental
verification is the Higgs particle responsible for electroweak
symmetry breaking.  While in the standard model the Higgs is
fundamental, and while almost everyone is firmly convinced that some
Higgs-like scalar field exits, there is a possibility that the
physical Higgs (like the scalar mesons) might itself be a bound state
of some deeper level of elementary particles (\eg, technicolor and its
variants). Despite the tremendous successes of the standard model of
particle physics we do not (currently) have direct proof of the
existence of a fundamental Higgs scalar field.

A third candidate scalar field of great phenomenological interest is
the axion: it is extremely difficult to see how one could make strong
interaction physics compatible with the observed lack of strong CP
violation, without something like an axion to solve the so-called
``strong CP problem''. Still, the axion has not yet been directly
observed experimentally.

A fourth candidate scalar field of phenomenological interest
specifically within the astrophysics/cosmology community is the
so-called ``inflaton''. This scalar field is used as a mechanism for
driving the anomalously fast expansion of the universe during the
inflationary era. While observationally it is a secure bet that
something like cosmological inflation (in the sense of anomalously
fast cosmological expansion) actually took place, and while scalar
fields of some type are the most reasonable way of driving inflation,
we must again admit that direct observational verification of the
existence of the inflaton field (and its variants, such as
quintessence) is far from being accomplished.

A fifth candidate scalar field of phenomenological interest
specifically within the general relativity community is the so-called
``Brans--Dicke scalar''. This is perhaps the simplest extension to
Einstein gravity that is not ruled out by experiment. (It is certainly
greatly constrained by observation and experiment, and there is no
positive experimental data guaranteeing its existence, but it is not
ruled out.) The relativity community views the Brans--Dicke scalar
mainly as an excellent testing ground for alternative ideas and as a
useful way of parameterizing possible deviations from Einstein
gravity. (And experimentally and observationally, Einstein gravity
still wins.)

Finally, the membrane-inspired field theories (low-energy limits of
what used to be called string theory) are literally infested with
scalar fields. In membrane theories it is impossible to avoid scalar
fields, with the most ubiquitous being the so-called ``dilaton''.
However, the dilaton field is far from unique, in general there is a
large class of so-called ``moduli'' fields, which are scalar fields
corresponding to the directions in which the background spacetime
geometry is particularly ``soft'' and easily deformed. So if membrane
theory really is the fundamental theory of quantum gravity, then the
existence of fundamental scalar fields is automatic, with the field
theory description of these fundamental scalars being valid at least
up to the Planck scale, and possibly higher.

So overall, while we have excellent theoretical reasons to expect that
scalar field theories are an integral part of reality, the direct
experimental/observational verification of the existence of
fundamental scalar fields is still an open question. Nevertheless, we
think it fair to say that there are excellent reasons for taking
scalar fields seriously, and excellent reasons for thinking that the
gravitational properties of scalar fields are of interest
cosmologically, astrophysically, and for providing fundamental probes
of general relativity.

\section{Problems with scalar field theories}
The main problem is that, generically, once you couple them to
gravity, they violate all the energy conditions even at a classical
level. We say generically because of the key role of the so-called
curvature coupling, a term of the form $\xi \phi^2 R$ in the
Lagrangian of the system that directly couples the scalar field $\phi$
with the spacetime curvature scalar $R$. This term is renormalizable
by power counting and so must be included in the curved-space scalar
field Lagrangian.  Even if this term is not there in the bare
Lagrangian it will be generated by quantum effects.

If $\xi=0$ (so-called ``minimal coupling'') then the SEC is violated
classically, though DEC, WEC, and NEC are satisfied. Unfortunately
``minimal coupling'' is non-generic and unstable to quantum
corrections. For any $\xi\neq0$ {\emph{all}} the pointwise energy
conditions are violated (including the NEC). There are good reasons to
believe that the value $\xi=1/6$ is preferred. Only for $\xi=1/6$
(so-called ``conformal coupling'') does the flat space limit of the
stress-energy tensor for the scalar field yield an expression with
good renormalization properties. This expression in flat space was
called the ``new improved stress-energy tensor'' to distinguish it
from the naive stress-energy tensor previously used \cite{new-improved}.

Indeed, from the quantum field theory perspective, conformal coupling
and the new improved stress-energy tensor are arguably the only
sensible choice, and it is rather disturbing that this choice leads to
violations of all pointwise energy condition (and so to peculiar
physics when coupled to gravity). Even worse, under certain
circumstances (typically involving trans-Planckian expectation values
for the scalar field) even the averaged null energy condition (ANEC)
is violated.  [Note that trans-Planckian values for a scalar field are
not by themselves objectionable; it is only trans-Planckian energy
densities that require a full quantum-gravity treatment. For example,
many (not all) models of cosmological inflation use trans-Planckian
values for the scalar field.]  The fact that the ANEC can be violated
by classical scalar fields is significant and important (even with the
trans-Planckian caveat).  The ANEC is the weakest of the energy
conditions in current use, and violating the ANEC short circuits {\em
all} the standard singularity/positive-mass/censorship theorems. This
observation piqued our interest and we decided to see just how weird
the physics could get once you admit scalar fields into your models.

In particular, it is by now well-known that traversable wormholes are
associated with violations of the NEC and ANEC, so we became
suspicious that there might be an explicit class of exact traversable
wormhole solutions to the coupled gravity-scalar field system. We
recently found such a class of [unstable]
solutions~\cite{wormholes,bronnikov}.  Now traversable wormholes,
while certainly exotic, are by themselves not enough to get the
physics community really upset: The big problem with traversable
wormholes is that if you manage to acquire even one inter-universe
traversable wormhole then it {\em seems} almost absurdly easy to build
a time machine --- and this does get the physics community upset.  At
this point, we will again confront ourselves with quantum physics.  It
has been conjectured by Hawking, ({\em Chronology Protection
Conjecture})~\cite{Hawking-cpc}, that quantum physics will save the
universe by destabilizing the wormhole just as a time machine is about
to form.  However, it must certainly be emphasized that there is
considerable uncertainty as to how serious these causality problems
are.

The violations of the energy conditions induced by non-minimally
coupled scalar fields, having a classical character, are not
restricted in magnitude or duration by any quantum
inequality~\cite{Ford-Roman-gsl}. Thus, even without reaching
trans-Planckian values for the scalar field, one can envisage the
creation of long-lasting fluxes of negative energy.  It is hard to see
how these negative energy fluxes can be made compatible with the
second law of thermodynamics~\cite{sl-violations}. As emphasized by
Ford and Roman~\cite{Ford-Roman-gsl}, the solution to this question is
tied up with the manner in which the energy flux interacts with
matter. In fact, trying to circumvent this issue by throwing the flux
into a black hole they found a miraculous preservation of the
generalized second law.

\section{Conclusions}

There are several responses to the current state of affairs: either we
can learn to live with wormholes, and other strange physics engendered
by energy condition violations, or we need to patch up the theory.
One particularly simple way of dealing with all these problems is to
banish scalar fields from your theories: This makes technicolor
partisans very happy, but drives supersymmetry supporters, string
theorists, and cosmologists to apoplexy.  Alternatively, one could
forbid non-minimal couplings, or forbid trans-Planckian field values,
each one of these particular possibilities is in conflict with
cherished notions of {\emph{some}} segments of the particle physics/
membrane theory/ relativity/ astrophysics communities.  Most
physicists would be loathe to give up the notion of a scalar field,
and conformal coupling is so natural that it is difficult to believe
that banning it would be a viable option. Banishing trans--Planckian
field values is more plausible, but this is only a partial remedy and
also runs afoul of at least some segments of the cosmological
inflationary community.

\bigskip

In summary: The conflict between quantum physics and gravity is now
becoming acute. Problems are no longer confined to Planck scale
physics but are leaking down to arbitrarily low energies and even into
the classical realm. These problems appear to be insensitive to and
independent of high energy phenomena and so it is not at all clear
that a high energy cutoff (string theory, quantum geometry, lattice
gravity, \etc\dots) would do anything to ameliorate them. The
situation is both puzzling and exciting.

\bigskip
\centerline{\#\ \#\ \#}
\bigskip



\begin{thebibliography}{99}



\bibitem{energy-conditions}
S.W. Hawking and G.F.R. Ellis, 
{\em The large scale structure of spacetime},
(Cambridge University Press, England, 1973). \\
R.M. Wald, 
{\em General Relativity},
(University of Chicago Press, Chicago, 1984). \\
M. Visser, 
{\em Lorentzian wormholes}, 
(AIP Press, New York, 1995).

\bibitem{consequences}
H. Bondi, Rev. Mod. Phys. {\bf 29}, 423 (1957). \\
M.S. Morris and K.S. Thorne, Am. J. Phys. {\bf 56}, 395 (1988). \\
M. Visser, Phys. Rev. D {\bf 39}, 3182 (1989); 
Nucl. Phys. {\bf B328}, 203--212
(1989). \\
M. Alcubierre, 
Class. Quantum Grav. {\bf 11}, L73 (1994) [gr-qc/0009013].

\bibitem{Science}
M. Visser,
Science {\bf 276}, 88 (1997).

\bibitem{Ford-Roman}
L.H. Ford and T.A. Roman,
Phys. Rev. D {\bf51}, 4277 (1995); D {\bf 60}, 104018 (1999); 
\\
L.H. Ford, M.J. Pfenning, and T.A. Roman,
Phys. Rev. D {\bf 57}, 4839 (1998). 

\bibitem{class-violations}
J.B. Bekenstein, Ann. Phys. (NY) 82, 535 (1974); Phys. Rev. D {\bf 11}, 2072 (1975). \\
S. Deser, Phys. Lett. B {\bf 134}, 419 (1984). \\
E.E. Flanagan and R.M. Wald, Phys. Rev. D {\bf 54}, 6233 (1996). 

\bibitem{surgery}
C. {\Barcelo} and M. Visser, Nucl. Phys. {\bf B584}, 415 (2000).
\bibitem{new-improved}
N.A. Chernikov and E.A. Tagirov, Ann. Inst. Henri Poincar\'e {\bf 9}, 109 (1968). \\
C.G. Callen, S. Coleman, and R. Jackiw, Ann. Phys. (NY) {\bf 59}, 72 (1970).

\bibitem{wormholes}
C. {\Barcelo} and M. Visser, 
Phys. Lett. {\bf B466}, 127 (1999);
Class. Quantum Grav. {\bf17}, 3843 (2000).

\bibitem{bronnikov}
K.~A.~Bronnikov and S.~Grinyok,
``Instability of wormholes with a nonminimally coupled scalar field,''
arXiv:gr-qc/0201083.

\bibitem{Hawking-cpc}
S.W. Hawking, Phys. Rev. D {\bf 46}, 603 (1992).

\bibitem{Ford-Roman-gsl}
L.H. Ford and T.A. Roman, Phys. Rev. D {\bf 64}, 024023 (2001). 

\bibitem{sl-violations}
L.H. Ford, Proc. Roy. Soc. Lond. A {\bf 364}, 227 (1978). \\
P.~C.~Davies and A.~C.~Ottewill,
``Detection of negative energy: 4-dimensional examples,''
arXiv:gr-qc/0203003.


\end{thebibliography}
\end{document}